\documentclass[twocolumn, superscriptaddress, nofootinbib, floatfix]{revtex4-2}

\usepackage{lmodern}
\usepackage{float}
\usepackage[colorlinks, allcolors=blue]{hyperref}
\usepackage{amsthm}
\usepackage{graphicx,caption}
\usepackage{tasks,amsthm,amsmath,amssymb,mathtools,parskip,enumerate,multirow,bbold,physics,mathtools}
\usepackage{tikz}
\usepackage[title]{appendix}
\usepackage[capitalise]{cleveref}

\usepackage{caption}

\theoremstyle{definition}

\newtheorem*{eg*}{Example~\ref{eg:simplest_nontrivial} cont}

\theoremstyle{remark}

\DeclareMathOperator{\pow}{pow}
\DeclareMathOperator{\powtwo}{pow2}
\DeclareMathOperator{\powthree}{pow3}
\DeclareMathOperator{\powfour}{pow4}
\DeclareMathOperator{\absolute}{abs}
\DeclareMathOperator{\negative}{neg}
\DeclareMathOperator{\mult}{mult}
\DeclareMathOperator{\division}{div}
\DeclareMathOperator{\add}{add}
\DeclareMathOperator{\subtract}{subtract}

\begin{document}

\title{Statistical Patterns in the Equations of Physics\\ and the Emergence of a Meta-Law of Nature} 

\author{Andrei Constantin}
\email{andrei.constantin@physics.ox.ac.uk}
\affiliation{School of Mathematics, University of Birmingham, Watson Building, Edgbaston, Birmingham B15 2TT, United Kingdom}
\affiliation{Rudolf Peierls Centre for Theoretical Physics, University of Oxford, Parks Road, Oxford OX1 3PU, UK}

\author{Deaglan Bartlett}
\email{deaglan.bartlett@iap.fr}
\affiliation{Rudolf Peierls Centre for Theoretical Physics, University of Oxford, Parks Road, Oxford OX1 3PU, UK}

\author{Harry Desmond}
\email{harry.desmond@port.ac.uk}
\affiliation{Institute of Cosmology and Gravitation, University of Portsmouth, Dennis Sciama
Building, Portsmouth, PO1 3FX, UK}

\author{Pedro G. Ferreira}
\email{pedro.ferreira@physics.ox.ac.uk}
\affiliation{Astrophysics, University of Oxford, DWB, Keble Road, Oxford OX1 3RH, UK}

\begin{abstract}
\noindent
Physics seeks to uncover the laws of Nature and express them through mathematical equations. Despite the vast diversity of natural phenomena, physical equations exhibit structural regularities that set them apart from arbitrary mathematical expressions. While principles such as dimensional analysis have long guided the formulation of physical models, the exploration of more subtle statistical patterns within the equations of physics remains  an open question. 
Here, by analysing four corpora of physics equations and applying advanced implicit-likelihood techniques, we find that the frequency of mathematical operators follows an exponential decay law, in contrast to Zipf's power law for word frequencies in natural languages.
This reveals a statistical meta-law of physics, possibly reflecting a combination of communication efficiency and constraints imposed by Nature itself. The meta-law offers practical benefits for symbolic regression by drastically narrowing down the space of physically plausible expressions. More broadly, it may inform the development of language models that can generate coherent mathematical representations, advancing the automation of physical law discovery. 
\end{abstract}
\maketitle

\section{Introduction}

The structure of physical reality is most precisely captured through mathematical equations~\cite{wigner1990unreasonable, feynman1967character, weinberg1994dreams}. These equations span a wide range of phenomena and include iconic examples such as Newton's law of gravitation, $F=GmM/r^2$~\cite{newton1833philosophiae}, Einstein's mass-energy equivalence,  $E=mc^2$~\cite{einstein1905does}, and the Hawking--Bekenstein formula for the entropy of black holes, $S=k_BAc^3/4G\hbar$~\cite{Bekenstein72, hawking1975particle}. When expressed in prefix notation, the right-hand sides of these equations resemble structured strings of text, where each ``word'' denotes either an operator (e.g., multiplication, division) or an operand (e.g., constants, variables):
\begin{equation}\label{example_expr}
\begin{aligned}
&{\rm div}({\rm mult}(G,{\rm mult}(m,M)),{\rm pow2}(r))~,\\
&{\rm mult}(m,{\rm pow2}(c))~,\\
&{\rm div}({\rm mult}(k_B,{\rm mult}(A,{\rm pow3}(c))),{\rm mult}(4,{\rm mult}(G,\hbar)))\,.
\end{aligned}
\end{equation}

The core structural features of physical equations can be traced to a handful of principles:
\begin{itemize}
\item[(i)] the presence of variables representing measurable quantities (e.g., the masses $m$ and $M$ or the horizon area $A$ in the formulae above);
\item[(ii)] the central role of universal constants such as $G$, $c$, $k_B$, and $\hbar$;
\item[(iii)] the frequent appearance of numerical coefficients;
\item[(iv)] constraints imposed by dimensional analysis, ensuring consistency of physical units; and
\item[(v)] the prevalence of power-law relationships and scaling behaviours, exemplified by the inverse-square law in Newtonian gravity.
\end{itemize}

These foundational features have long served as guiding principles for physicists when modeling natural phenomena. Yet beyond these well-known structural elements, it remains unclear whether physical equations exhibit more subtle, statistical regularities.
To address this question, here we analyse the distribution of mathematical operators and operands across four corpora of physics equations: one extracted from \textit{The Feynman Lectures on Physics}~\cite{Feynman:1494701, udrescu2020ai}; one derived from the subset of physics equations in Wikipedia's \textit{List of Scientific Equations Named after People}\cite{guimera2020bayesian, SRPriors_2023, PhysRevD.109.083524}; one based on the comprehensive \textit{Encyclopaedia Inflationaris} review of inflationary cosmology~\cite{martin2014encyclopaedia}; and one compiled from \textit{The Cambridge Handbook of Physics Formulas}~\cite{woan2000cambridge}.

This investigation draws inspiration from linguistics, where \textbf{Zipf’s law} describes the frequency distribution of words in natural languages~\cite{estoup1916gammes, dewey1923relativ, condon1928statistics, kingsley1932selected}. Specifically, it states that the frequency $f$ of a word is inversely proportional to its rank $r$ in the frequency table, typically following a power law $f(r) \propto r^{-\alpha}$ with $\alpha \approx 1$, largely independent of language, genre, or time period:
\begin{equation}
\label{Zipf_law}
    \text{Zipf's law:}~~~ f(r)\sim \frac{1}{r^\alpha}~.
\end{equation}
This implies that the most frequent word appears roughly twice as often as the second most frequent word, three times as often as the third, and so on.

Despite its ubiquity, the origin of Zipf’s law remains debated. Proposed explanations span a range of theories, including communication efficiency~\cite{zipf1935psycho, zipf1949human, piantadosi2011word, kanwal2017zipf}, stochastic models of text generation~\cite{simon1955class, zanette2005dynamics}, and mechanisms involving history-dependent reductions in sample space~\cite{thurner2015understanding}.
Zipf’s law and its two-parameter generalisation, 
\begin{equation}
\label{Zipf_Mandelbrot_law}
    \text{the Zipf--Mandelbrot law:}~~~ f(r)\sim \frac{1}{(r+b)^\alpha}~,
\end{equation}
have been observed in a wide variety of domains beyond language. These include the distribution of city sizes~\cite{makse1995modelling, gabaix1999zipf}, website popularity and file sizes in computer systems~\cite{breslau1999web, blank2000power}, as well as wealth inequality and the popularity of given names~\cite{cristelli2012there, zanette2001vertical}.
Zipf's law even extends to neuroscience: the firing rates of neurons in the brain exhibit a Zipfian distribution, with a few neurons highly active while most remain comparatively inactive~\cite{mora2011biological, tyrcha2013effect, schwab2014zipf}. In genomics, the abundance of expressed genes similarly follows a Zipfian pattern, indicating that only a small subset of genes is highly expressed while the majority remain at low expression levels~\cite{furusawa2003zipf}. Remarkably, Zipf's law has also been observed in astrophysics, where it describes the size distribution of galaxy superclusters~\cite{de2021zipf}.

Zipf's law has also been observed in programming languages, where it governs the frequency distribution of language elements such as keywords (e.g., \texttt{if}, \texttt{for}, \texttt{while}), operators (e.g., $+$, $-$, $\ast$, $=$), and developer-defined identifiers (e.g., variable and function names)~\cite{ellis1986emergence, hindle2016naturalness, allamanis2013mining, 10043269, casalnuovo2019studying}. These Zipfian patterns have informed the development of efficient tools for code prediction and completion, enhancing the performance of language models in programming contexts. Interestingly, an early study by Ellis and Hitchcock~\cite{ellis1986emergence} found that the command usage patterns of experienced programmers adhered more closely to Zipf's law than those of novices, suggesting that conformity to the distribution could serve as an indicator of efficient command use.

The present study focuses on the distribution of operators in mathematical expressions that encode physical laws. Mathematics is often described as a language, with symbols and operators conveying information about quantities and their relationships, much like words convey ideas in natural languages. From this perspective, it is natural to ask whether statistical patterns analogous to Zipf's law might also emerge in mathematical corpora. For example, one could analyse the symbol usage in Whitehead and Russell’s three-volume \textit{Principia Mathematica}~\cite{russell25}, or study frequency distributions across the \textit{mathlib} Lean library of digitised mathematics, which currently contains over 300{,}000 theorems and definitions~\cite{wikipedia_lean}. 

However, our present aim is more specific. Physics equations capture in closed-form expressions the fundamental patterns that govern natural phenomena. As such, the statistical regularities we study may not only reflect the mathematical structure and logical consistency required for effective scientific communication, but also hint at deeper, intrinsic constraints imposed by Nature itself.

Much like Zipf's law, whose origins remain debated but whose utility is widely recognized, uncovering statistical patterns in physics equations may yield significant practical benefits. In particular, such patterns can enhance symbolic regression algorithms by drastically reducing the search space and filtering out non-physical candidates.
A similar idea has already been explored in Ref.~\cite{guimera2020bayesian}, where prior probabilities over mathematical expressions were computed from the empirical frequencies of operators in a formula corpus extracted from Wikipedia. Subsequent work~\cite{SRPriors_2023, PhysRevD.109.083524} extended this by training language models that also capture correlations between operators, further refining these priors. Our findings suggest that such statistical structures are not arbitrary but may reflect a deeper ``meta-law'' governing the form of physical equations. Incorporating this meta-law into symbolic regression and scientific language models could significantly enhance their capacity for automated hypothesis generation, guiding the discovery of new, physically meaningful mathematical expressions.

\section{The physics corpora}

\begin{figure*}[t] 
	\centering
	\includegraphics[width=.9\textwidth]{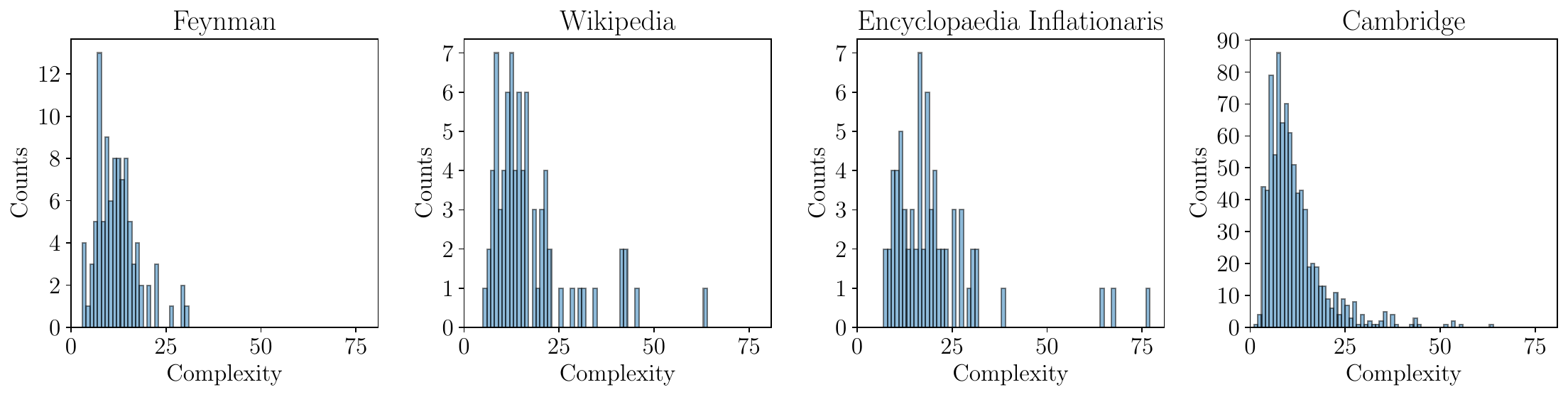} 
		\caption{{\bfseries Distribution of expression complexity (number of symbols appearing in an equation) in the four corpora.}	}
	\label{ComplexityAll} 
\end{figure*}

We analyse the distribution of operators in physical equations of the form
\begin{equation}
\text{quantity} = \text{closed-form expression}.
\end{equation}
Our study focuses on the mathematical expressions on the right-hand side, which we assume to involve only nullary, unary, and binary operators. Importantly, we exclude dynamical laws that contain differential or integral operators. In other words, we restrict our attention to physical laws expressible in closed-form expressions, a choice particularly well suited to identifying statistical patterns relevant for symbolic regression.

We distinguish three types of \emph{nullary} operators:
\begin{itemize}
    \item[(i)] \textbf{Variables}, denoted generically by the symbol~$x$;
    \item[(ii)] \textbf{Numerical constants}, denoted by~$a$;
    \item[(iii)] \textbf{Physical or mathematical constants}, denoted by~$c$.
\end{itemize}
For example, in the Hawking--Bekenstein entropy formula, the right-hand side includes one variable (the black hole horizon area $A$), one numerical factor ($4$), and four physical constants ($k_B$, $c$, $G$, and $\hbar$). We group both physical constants (e.g., $G$) and mathematical constants (e.g., $\pi$, Apéry’s constant $\zeta(3)$) in the same category due to their foundational roles in physics. While one might consider distinguishing between dimensional and dimensionless constants, such a classification would, for instance, place the fine-structure constant in the same category as ordinary numbers, which we find less intuitive.

The set of \emph{unary} operators frequently encountered in physics includes: $\absolute$, $\exp$, $\log$, $\negative$, $\sqrt{\cdot}$, as well as the (hyperbolic) trigonometric functions and their inverses. Our notation is such that $\absolute(x) = |x|$, $\negative(x) = -x$, and all other operators carry their standard mathematical meaning. We additionally distinguish the square, cube, and fourth power functions, written as $\powtwo(x) = x^2$, $\powthree(x) = x^3$, and $\powfour(x) = x^4$, since small integer exponents occur with disproportionately high frequencies in physical equations. While this categorisation is ultimately a convention, its influence on the overall operator distribution is minimal.

Lastly, we identify five principal \emph{binary} operators: $\mult$, $\division$, $\add$, $\subtract$, and $\pow$, with notation such as $\subtract(x, y) = x - y$ and $\pow(x, y) = x^y$. We also account for less common binary operations that appear in specialised contexts, such as spherical harmonics.

We analyse four distinct corpora of physical formulae.
The first corpus consists of 100 equations extracted from {\itshape The Feynman Lectures on Physics}, previously curated in Ref.~\cite{udrescu2020ai} to evaluate the AI Feynman symbolic regression model. These expressions span foundational topics in classical mechanics, electrodynamics, thermodynamics, statistical mechanics, and quantum mechanics. In total, the dataset comprises 1,173 operator instances drawn from 19 distinct operator types, with expression lengths ranging from 3 to 30 operators (\cref{ComplexityAll}).

The second corpus includes 78 expressions sourced from physics-related entries in Wikipedia’s {\itshape List of Scientific Equations Named after People}. A subset of this collection was employed in Refs.~\cite{SRPriors_2023,PhysRevD.109.083524} to train language models aimed at estimating prior probabilities of symbolic expressions. The dataset spans 839 operator instances across 19 unique types, with expression lengths ranging from 3 to 62 operators (\cref{ComplexityAll}). It covers a broad spectrum of physical subfields, including acoustics, astrophysics, biophysics, cosmology, earth sciences, electromagnetism, fluid dynamics, materials science, optics, physical chemistry, relativity, solid-state physics, and thermodynamics. 

The third corpus comprises 71 distinct expressions for the potential energy density of the inflaton field, a central concept in early universe cosmology. These expressions are drawn from the comprehensive review {\itshape Encyclopaedia Inflationaris}~\cite{martin2014encyclopaedia}, and together account for 1,371 operators spanning 23 operator types. Expression lengths range from 7 to 77 operators (\cref{ComplexityAll}).

The fourth and largest corpus is derived from {\itshape The Cambridge Handbook of Physics Formulas}~\cite{woan2000cambridge}, a comprehensive compendium of closed-form expressions encompassing nearly all areas of physics. It includes $852$ expressions with lengths varying from 2 to 62 operators and a total of 9,449 operator instances across 40 unique types. The handbook covers diverse domains including dynamics, quantum mechanics, thermodynamics, solid-state physics, electromagnetism, optics, and astrophysics.

For each corpus, we determine the frequency of each operator type by counting its total number of occurrences and normalising by the total number of operators in the corpus. This yields a probability distribution over operator types. The operators are then ranked in descending order of frequency, with rank 1 corresponding to the most common operator, rank 2 to the second most common, and so on.
This rank–frequency relationship allows us to investigate whether the distribution follows a power-law, exponential, or other functional form, analogous to Zipf’s law in natural language. Our aim is to identify statistical regularities that may reflect universal structural features of physical laws. These patterns are then contrasted with those found in randomly generated symbolic expressions (see Section~\ref{sec:binary trees}), which serve as a null model for evaluating deviations from randomness in real physical formulae.

\section{Random mathematical expressions}
\label{sec:binary trees}

Before presenting our results, it is instructive to consider the baseline expectation for operator frequency distributions in corpora composed of randomly generated mathematical expressions with sizes and complexities comparable to those studied here. Such expressions, built from nullary, unary, and binary operators, can be naturally represented as binary trees, where each node has zero, one, or two children. For instance, Newton’s law of universal gravitation from \cref{example_expr} can be visualised as the binary expression tree shown in Fig.~\ref{fig:tree_example}.

\begin{figure}[h]
    \centering
     \begin{tikzpicture}[
        level distance=1cm,
        level 1/.style={sibling distance=2.8cm},
        level 2/.style={sibling distance=1.7cm},
        every node/.style={circle, draw, minimum size=2.4em, inner sep=0pt, font=\small}
    ]
        \node {{\scriptsize div}}
            child { node {{\scriptsize mult}}
            child { node {$G$} }
            child { node {{\scriptsize mult}}
                child { node {$m$} }
                child { node {$M$} }
            }
         }
            child { node {{\scriptsize pow2}}
            child { node {$r$} }
            };
    \end{tikzpicture}
    \caption{{\bfseries Binary expression tree for Newton's law of gravitation.}
    }
    \label{fig:tree_example}
\end{figure}

In this representation, external nodes (i.e., nodes with no children) correspond to nullary operators, such as variables, constants, or numerical coefficients. Internal nodes correspond to either unary or binary operators, depending on whether they have one or two children, respectively. This structure leads to a simple but important combinatorial result: any binary expression tree with $i$ binary operators contains exactly $i+1$ nullary operators, regardless of the number $j$ of unary operators. The total number of operators in the expression is then given by
\begin{equation}\label{length_formula}
n = 2i + j + 1,
\end{equation}
which we adopt as a measure of the expression's complexity (or length). 

The universal relationship between the number of nullary and binary operators has significant implications for the expected operator frequency distribution, especially in the limit of large expression complexity, $n \gg 1$. In any collection of expressions with a typical number of binary operators $i \gg 1$, the relative frequencies of the three operator classes---nullary, unary, and binary---are approximately given by:
\begin{equation}
f_0 : f_1 : f_2 \simeq \frac{i}{n} : 1 - \frac{2i}{n} : \frac{i}{n}~.
\end{equation}
This expression follows directly from Eq.~\eqref{length_formula} in the limit $i,n\gg 1$. 
Naturally, the specific value of the ratio $i/n$ depends on the structural characteristics of a given corpus. For instance, in a collection of high-complexity expressions with no unary operators ($j=0$), the distribution simplifies to 
\[
f_0 : f_1 : f_2 = \frac{1}{2} : 0 : \frac{1}{2}~.
\]
At the opposite extreme, in corpora composed of randomly generated symbolic expressions with balanced use of all operator arities, the frequency distribution tends toward
\begin{equation} \label{unif_distr}
f_0 : f_1 : f_2 = \frac{1}{3} : \frac{1}{3} : \frac{1}{3}~,
\end{equation}
a result that can be derived using elementary combinatorial arguments, as follows. 

\subsection{Asymptotic operator distributions in random expressions}

Random unlabeled binary expression trees can be generated by initialising a single active node (the root) and recursively assigning a number of children to each active node. Assigning two children increases the number of active nodes by one, assigning one child leaves it unchanged, and assigning zero children (i.e., creating a leaf) decreases the number by one. Continuing this process until no active nodes remain maps the tree-building procedure to a one-dimensional random walk with $n$ steps: the walk begins at $1$ (one active node) and ends at $0$ (no active nodes), with each step representing a node in the expression tree and corresponding to a step of $+1$ (binary operator), $0$ (unary operator), or $-1$ (nullary operator).

For example, if nodes are processed in a left-to-right, top-to-bottom order, the binary tree shown in Fig.~\ref{fig:tree_example} maps to the sequence of active node counts
\begin{equation*}
1 \rightarrow 2 \rightarrow 3 \rightarrow 3 \rightarrow 2 \rightarrow 3 \rightarrow 2 \rightarrow 1 \rightarrow 0,
\end{equation*}
which corresponds to the $n=8$ step walk
\begin{equation*}
(1, 1, 0, -1, 1, -1, -1, -1),    
\end{equation*}
where each entry indicates the change in the number of active nodes after visiting a node: $+1$ for a binary operator (two children), $0$ for a unary operator (one child), and $-1$ for a nullary operator (no children). 

The total number of such walks involving $n$ steps, where $i$ nodes are binary ($+1$ steps), $i{+}1$ are nullary ($-1$ steps), and the remaining $n{-}2i{-}1$ are unary ($0$ steps), is given by the multinomial coefficient
\begin{equation}\label{binomial}
\frac{n!}{i!(i+1)! (n - 2i - 1)!}~,
\end{equation}
since this corresponds to the number of permutations of the $n$-tuple
\begin{equation}
(\underbrace{1, \ldots, 1}_{i\text{ times}},\ 
\underbrace{0, \ldots, 0}_{n - 2i - 1\text{ times}},\ 
\underbrace{-1, \ldots, -1}_{i+1\text{ times}}).
\end{equation}

Not all of these paths correspond to valid connected binary trees: those that return to zero before the final step describe disjoint or incomplete structures. However, there exists an $n$-to-$1$ mapping from the full set of such walks to the subset that first returns to zero at the final step (i.e., proper expression trees).

Using this, the expected fraction of binary operators, $\langle i/n \rangle$, can be computed exactly by weighting each configuration by the number of corresponding walks:
\begin{equation}\label{expectation_value}
\left\langle \frac{i}{n} \right\rangle 
= \frac{\displaystyle \sum_{i=0}^{\left\lfloor\frac{n-1}{2}\right\rfloor} \frac{(n-1)!}{i!(i+1)! (n - 2i - 1)!} \cdot \frac{i}{n}}
       {\displaystyle \sum_{i=0}^{\left\lfloor\frac{n-1}{2}\right\rfloor} \frac{(n-1)!}{i!(i+1)! (n - 2i - 1)!}}~,
\end{equation}
where $\lfloor \cdot \rfloor$ denotes the floor function.

An alternative approach for enumerating such trees involves Catalan numbers. The $i^\text{th}$ Catalan number $C_i$ counts the number of Dyck paths of length $2i$ that never cross below zero, i.e., sequences of $i$ $(1)$'s and $i$ $(-1)$'s with non-negative partial sums. All valid expression trees of length $n$ can be generated by interleaving $n - 2i - 1$ unary nodes (represented by $0$'s) into such Dyck paths, and appending a final $-1$ to ensure the walk terminates properly. See Ref.~\cite{lample2019deep} for a related discussion.

In the large-$n$ limit, the sum in Eq.~\eqref{expectation_value} is dominated by terms satisfying $i \approx i + 1 \approx n - 2i - 1$, i.e., where each of the three node types appears with roughly equal frequency. This leads directly to the uniform distribution
\begin{equation*}
f_0 : f_1 : f_2 = \frac{1}{3} : \frac{1}{3} : \frac{1}{3}~,
\end{equation*}
as given in Eq.~\eqref{unif_distr}, confirming the combinatorial origin of the expected frequency profile in randomly generated symbolic expressions.

\subsection{Corpora of random mathematical expressions}
Under this assumption of uniformity across arity classes, and assuming a flat distribution within each class, a corpus containing $n_0$ distinct nullary operators, $n_1$ unary operators, and $n_2$ binary operators is expected to exhibit the following per-operator frequency:
\begin{equation} \label{flat_distribution}
\frac{1}{3n_0},\quad \frac{1}{3n_1},\quad \frac{1}{3n_2},
\end{equation}
corresponding to the nullary, unary, and binary categories, respectively.

\begin{figure}
    \centering
    \includegraphics[width=0.4\textwidth]{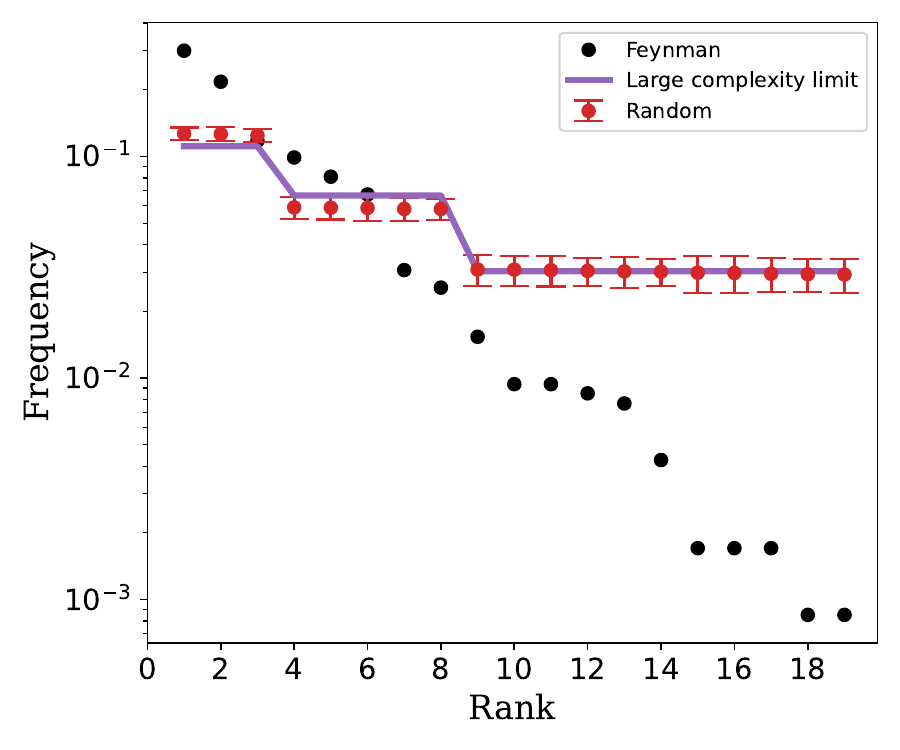}
\caption{{\bfseries Comparison of operator frequency distributions between the {\itshape Feynman} corpus and randomly generated expression corpora.}
Black points represent the operator frequencies observed in the {\itshape Feynman} corpus (see~\cref{table:alldata}). Red points show the mean frequencies and standard deviations computed over 100 randomly generated corpora matched in size and complexity distribution to the {\itshape Feynman} corpus. The purple line indicates the expected piecewise-uniform distribution in the large-complexity limit, as predicted by~\cref{flat_distribution}.
}     
    \label{StatisticalSignificance}
\end{figure}

To empirically illustrate this expectation, we generated 100 mock corpora of random mathematical expressions, each matched in size and complexity distribution to the Feynman corpus. The generation process involves two steps: first, a random binary tree structure is sampled (equivalent to a one-dimensional Dyck path, as described above); second, operators are assigned to the nodes uniformly at random from their respective arity classes. The resulting average frequencies and standard deviations are presented in \cref{StatisticalSignificance}. As expected, the distributions are nearly piecewise flat, closely matching the predictions of Eq.~\eqref{flat_distribution} and reflecting the assumption that operators of the same arity are drawn from uniform distributions.

\section{Statistical patterns in physics formulae}
\label{sec:analysis}

 {
\renewcommand{\arraystretch}{1.1}
\begin{table*}
\centering
\caption{\textbf{
Ranked operator frequencies in four corpora of physics equations.}}
\label{table:alldata}
\begin{tabular}{|c|c|}
    \hline
{\bfseries Corpus} & {\bfseries (rank $r$, frequency $f$, operator) } \\\hline\hline
    \parbox{1.5cm}{\begin{equation*}\begin{aligned}\text{Feynman's} \\ \text{Lectures~~}\\ \text{on Physics}\end{aligned}\end{equation*}} &
    \parbox{11.cm}{
    \begin{equation*}
        \begin{array}{l l l l }
            (1, 0.30, x) &  (6, 0.067, \powtwo) &  (11, 0.0094,\sin)  &(16, 0.0017, \powfour )    \\
            (2, 0.22, \mult) &  (7, 0.031, \subtract) & (12, 0.0085, \exp ) & (17, 0.0017, \tan)  \\
            (3, 0.12, \division) & (8, 0.026, \add ) & (13, 0.0077, \negative) & (18, 0.00085, \pow )   \\
            (4, 0.099, a) &  (9, 0.015, \sqrt{\cdot})   & (14, 0.0043, \powthree)   &  (19, 0.00085, \tanh )    \\
             (5, 0.083, c) &(10, 0.0094, \cos)  &(15, 0.0017, \arcsin ) &
        \end{array}
    \end{equation*}
    }
  \\\hline
    \parbox{1.5cm}{\begin{equation*}\centering \begin{aligned} \text{Wikipedia} \\ \text{named~~ }\\ \text{equations}\end{aligned}\end{equation*}}&
    \parbox{11cm}{
    \begin{equation*}
        \begin{array}{l l l l }
            (1, 0.34, x) &  (6, 0.045, \subtract) &  (11, 0.0083, \exp)  &  (16, 0.0024, {\rm arcsinh})\\
            (2, 0.16, \division) &  (7, 0.042, c)  &  (12, 0.0083, \log)    & (17, 0.0024, \sin) \\
            (3, 0.15, \mult) & (8, 0.035, \powtwo) &  (13, 0.0072, \cos)  &   (18, 0.0012, \sinh) \\
             (4, 0.10, a)  & (9, 0.024,\pow) & (14, 0.0048, \powthree) & (19, 0.0012, \tanh) \\
             (5, 0.049, \add) & (10, 0.0095,\negative) & (15, 0.0048, \sqrt{\cdot}) &
        \end{array}
    \end{equation*}
    }
    \\\hline
    \parbox{1.5cm}{\begin{equation*}\begin{aligned}\text{Encycl.~~~\,} \\ \text{\!\!Inflationaris}\end{aligned}\end{equation*}}&
    \parbox{11cm}{
    \begin{equation*}
        \begin{array}{l l l l  }
            (1, 0.26, x)  & (7, 0.043, \subtract)   & (13, 0.016, \negative )  & (19, 0.0014, \arctan)    \\
            (2, 0.16, \mult ) &  (8, 0.041, \add ) &(14, 0.012, \log)  & (20, 0.0014, \powthree)   \\
            (3, 0.12, a) &(9, 0.033, \powtwo)   & (15, 0.0065, \cos)  &(21, 0.00072, \absolute )    \\
            (4, 0.12, \division) & (10, 0.022, \pow)  & (16, 0.0051, \tan )  & (22, 0.00072, \cosh)   \\
            (5, 0.059, \powfour) &   (11, 0.019, \exp) & (17, 0.0043,\sin )  & (23, 0.00072, \sinh ) \\
             (6, 0.051, c)  &  (12, 0.018, \sqrt{\cdot} )  & (18, 0.0022, \tanh ) & 
        \end{array}
    \end{equation*}
    }
    \\\hline
 \parbox{1.5cm}{\begin{equation*}\begin{aligned}\text{Cambridge} \\ \text{Handbook\,}\\ \text{of Physics~}\\ \text{Formulas~~}\end{aligned}\end{equation*}}&
    \parbox{10cm}{
   \begin{equation*}
    \begin{array}{l l l}
        (1, 0.27, x) & (12, 0.0079, \cos) & (23, 0.00042, \tanh) \\
        (2, 0.22, \mult) & (13, 0.0075, \pow) & (24, 0.00032, {\rm sinc}) \\
        (3, 0.13, a) & (14, 0.0073, \exp) & (25-29, 0.00021, \{\sinh, \cosh, \\
        (4, 0.10, \division) & (15, 0.0065, \powthree) & ~~~~~~~~~~\coth,{\rm Langevin},\arcsin\})\\
        (5, 0.076, c) & (16, 0.0035, \powfour) &  (30-40, 0.00011, \{\text{arccosh}, \\
        (6, 0.058, \powtwo) & (17, 0.0034, \log) & ~~~{\rm arcsinh}, {\rm Debye},{\rm Hermite},\\
        (7, 0.030, \add) & (18, 0.0013, \absolute) & ~~~{\rm Bessel}, \sec, {\rm Airy}, {\rm spectral~fn.},\\
        (8, 0.030, \subtract) & (19, 0.0011, \log_{10}) &~~~{\rm Brillouin}, {\rm Sph.~harmonic},\\
        (9, 0.020, \sqrt{\cdot}) & (20, 0.00085, \tan) & ~~~ {\rm Legendre}\}) \\
        (10, 0.014, \negative) & (21, 0.00074, \text{Cornu}) &  \\
        (11, 0.0081, \sin) & (22, 0.00042, \csc) & 
    \end{array}
\end{equation*} 
    }
    \\\hline    
\end{tabular}
\end{table*}
}

The operator frequency distributions for the three corpora are summarised in \cref{table:alldata} and show a marked deviation from the flat, piecewise-uniform distribution predicted by our random generation model (\cref{flat_distribution}). As illustrated in \cref{StatisticalSignificance}, even in the case of the relatively small {\itshape Feynman} corpus, the observed operator frequencies exhibit a statistically significant structure.

Strikingly, the operator frequency distributions across all three corpora display a high degree of similarity, which raises a natural question: can these distributions be approximated by a simple, universal law, akin to frequency--rank relations commonly observed in linguistics?

To explore this, we considered three candidate functional forms for the frequency--rank relationship. Motivated by linguistic analogies, we examined both Zipf's law from \cref{Zipf_law} and the more general Zipf--Mandelbrot law from \cref{Zipf_Mandelbrot_law}, as well as an exponential model:
\begin{equation}
    \label{Exponential_law}
    f(r) \sim \exp(-\beta r),
\end{equation}
where $\beta > 0$ governs the rate of exponential decay.

Fitting these distributions poses a challenge, as the frequency--rank data points are inherently correlated: by construction, the frequencies must decrease monotonically with increasing rank. Traditional methods such as least-squares or Poisson likelihoods fail to account for these correlations adequately. We therefore adopt a simulation-based inference approach~\cite{Cranmer_2020,marin2012approximate}, as detailed below.

\subsection{Fitting a law to operator frequencies}
\label{sec:fitting_law}

\begin{table*}[t]
    \centering
\caption{\textbf{Inferred parameters for the three frequency--rank distributions} (Zipf’s law, Zipf--Mandelbrot law, and exponential law; see \cref{Zipf_law,Zipf_Mandelbrot_law,Exponential_law}) across the corpora studied in this work. Also shown are the logarithms of the Bayesian evidence ratios, $\log_{10} K$, computed relative to the exponential model. Positive values indicate a statistical preference for the exponential distribution, which is consistently favored across all corpora.}
    \label{tab:parameters}

    \begin{tabular}{c|c|c|c|c}
         & Feynman & Wikipedia & Encyclopaedia Inflationaris & Cambridge \\
         \hline\hline
         Zipf & $\alpha = 2.6 \pm 0.8$ & $\alpha = 2.3 \pm 0.3$ & $\alpha = 2.6 \pm 0.5$ & $\alpha = 1.7 \pm 0.3$\\
         & $\log_{10} K = 12.2$ & $\log_{10} K = 11.2$ & $\log_{10} K = 10.2$ & $\log_{10} K = 11.9$ \\[2pt]
         \hline
         Zipf--Mandelbrot & $\alpha = 5.9 \pm 1.3$ & $\alpha = 11.5 \pm 4.1$ & $\alpha = 5.9 \pm 1.2$ & $\alpha = 8.8 \pm 0.9$ \\
          & $b = 15.7 \pm 2.9 $ & $b = 16.5 \pm 2.6$ & $b = 14.0 \pm 3.8$ & $b = 13.6 \pm 2.4$ \\
          & $\log_{10} K = 12.2$ & $\log_{10} K = 2.9$ & $\log_{10} K = 2.8$ & $\log_{10} K = 11.2$ \\[2pt]
          \hline
         Exponential & $\beta = 0.342 \pm 0.010$ & $\beta = 0.35 \pm 0.03$ &  $\beta = 0.266 \pm 0.014$ &  $\beta = 0.33 \pm 0.04$ \\
    \end{tabular}
    
\end{table*}

Although the likelihood for the operator frequency--rank relation is analytically intractable, generating synthetic samples from candidate distributions is straightforward. We therefore employ simulation-based inference (also known as likelihood-free or implicit likelihood inference) to perform our analysis~\cite{Cranmer_2020,marin2012approximate}.
For each distribution -- Zipf law (with exponent $\alpha$), the Zipf--Mandelbrot law (with parameters $\alpha$ and~$b$), and the exponential law (with decay rate $\beta$) -- we perform 10,000 simulations. In each simulation, parameters are sampled from the following prior ranges:
\begin{equation}
\alpha \in [1, 10],\quad b \in [2, 20],\quad \beta \in [0.1, 1],    
\end{equation}
with uniform priors applied to $\alpha^{-1}$, $b^{-1}$, and $\beta$. The resulting parameters define a theoretical frequency--rank distribution, from which we draw $N$ samples (where $N$ corresponds to the number of observed operators in the target corpus). These samples are re-binned and sorted to construct a synthetic frequency--rank curve.

To enable consistent inference across simulations with different numbers of unique operators, we compress each synthetic dataset into a fixed-length feature vector. Specifically, we fit each synthetic frequency--rank curve to the three candidate models (\cref{Zipf_law,Zipf_Mandelbrot_law,Exponential_law}), allowing for an overall normalisation parameter. From each fit, we record the optimised parameters and the mean squared error (10 numbers). We also keep track of the total number of operators, yielding a summary vector of 11 features.

We then train neural posterior estimators using the \textsc{ltu-ili} package~\cite{Ho_2024}, to learn the conditional distribution of model parameters given these 11 summary features. Our ensemble consists of eight neural networks: four mixture density networks and four masked autoregressive flows, each using 3--6 coupling layers and 25 or 50 hidden units. Training is conducted with a batch size of 64 and a learning rate of $10^{-4}$, using 80\% of simulations for training and 20\% for validation. Model calibration is verified through standard metrics, including percentile--percentile tests and the `Tests of Accuracy with Random Points' diagnostic~\cite{Lemos_2023}.

\begin{figure*} 
	\centering
	\includegraphics[width=\textwidth]{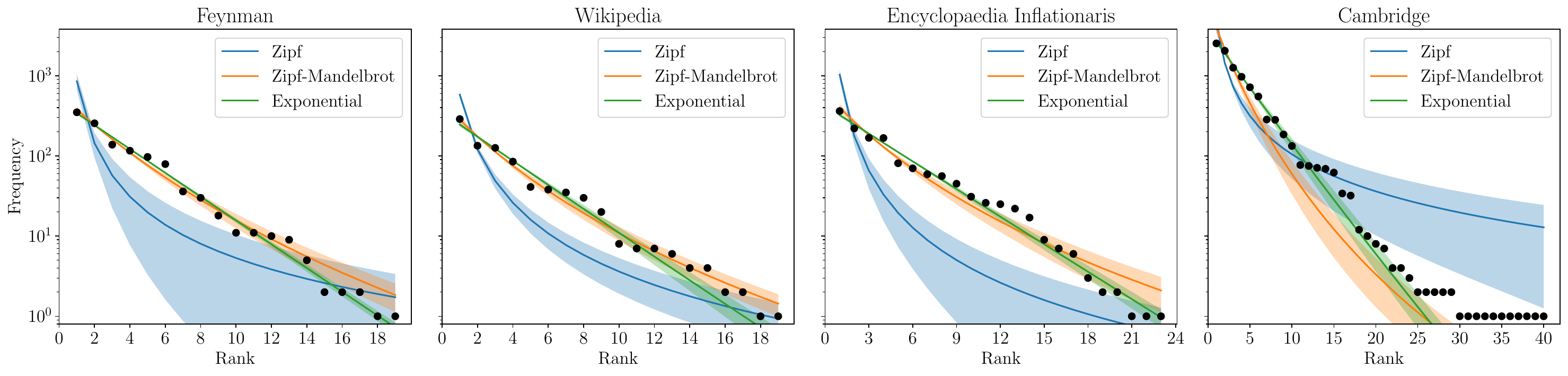} 
\caption{\textbf{Posterior distributions of the fits across corpora.}
We compare Zipf (\cref{Zipf_law}), Zipf--Mandelbrot (\cref{Zipf_Mandelbrot_law}), and exponential (\cref{Exponential_law}) models. The exponential fit clearly provides the best agreement with the data. Solid lines show the posterior mean; shaded regions indicate 68\% confidence intervals.}
	\label{fig:ili_fits} 
\end{figure*}

Once trained, the posterior estimators are evaluated on the real corpora. The compressed summary vectors from each corpus are passed through the networks to obtain posterior distributions over the parameters of each model. The resulting parameter constraints are reported in \cref{tab:parameters}. From the posterior samples, we compute the expected frequency--rank curves and 68\% confidence intervals, which are plotted alongside the empirical data in \cref{fig:ili_fits}.

As evident in \cref{fig:ili_fits}, the Zipf law provides a visibly poor fit across all corpora. In contrast, both the exponential and Zipf--Mandelbrot models provide good visual agreement with the data. To quantitatively assess model preference, we compute Bayesian evidence ratios~$K$ using evidence networks~\cite{Jeffrey_2024}, as implemented in \textsc{ltu-ili}. These are trained using two-layer neural networks with 64 hidden units per layer, a batch size of 512, and a learning rate of $10^{-9}$ for 100 epochs.

The evidence ratios are summarised in \cref{tab:parameters}. As anticipated, the Zipf model is decisively ruled out across all corpora: the lowest value of $\log_{10} K$ compared to the exponential model exceeds 10.2. According to the Jeffreys scale~\cite{Jeffreys_1998}, values of $\log_{10} K > 2$ constitute decisive evidence in favour of the preferred model.

While the Zipf--Mandelbrot and exponential models produce comparable fits by eye, the exponential law is consistently favoured in all four corpora, albeit with smaller evidence ratios in some cases. The {\itshape Feynman} and {\itshape Cambridge} corpora yield the strongest preference for the exponential model, with $\log_{10} K = 12.2$ and $11.2$, respectively. The {\itshape Wikipedia} and {\itshape Encyclopaedia Inflationaris} corpora also favour the exponential law, with $\log_{10} K = 2.9$ and $2.8$, respectively.
This preference is not only statistically supported but also simpler: the exponential law requires one fewer parameter than the Zipf--Mandelbrot law. Furthermore, all four corpora yield consistent estimates of the exponential decay rate, with $\beta \sim 0.3$, suggesting the existence of a universal structural principle governing operator usage in mathematical expressions across physics.

\subsection{Results}

The findings, summarised in \cref{fig:ili_fits}, highlight four main observations:
\begin{itemize}
  \item[(i)] Operator distributions in physical equations deviate from Zipf’s law, lacking the power-law decay typical of natural language.
  \item[(ii)] Across all four corpora, the data are well described by an exponential distribution of the form~\eqref{Exponential_law}, with a consistent decay rate of $\beta \approx 0.3$.
  \item[(iii)] Variables, constants (numerical and physical), multiplication, and division dominate the top ranks, suggesting a structural preference in the formulation of physical laws.
  \item[(iv)] Unary operators are rare, accounting for no more than 1/5 of all operators in the {\itshape Encyclopaedia Inflationaris}, and under 1/7 in the other corpora.
\end{itemize}

The {\itshape Feynman}, {\itshape Wikipedia}, and {\itshape Cambridge} corpora all follow the same frequency--rank relation:
\begin{equation}
    \label{eq:exponential_3}
    f(r) \sim e^{-r/3},
\end{equation}
despite differences in operator ordering. It is worth noting that the {\itshape Cambridge} corpus, which contains nearly 10{,}000 operators, includes a long tail of low-frequency operators that appear only once or twice. These rare operators have large relative uncertainties and thus contribute little to the fit -- analogous to a Poisson likelihood, where the relative error scales as $1/\sqrt{N}$. While their statistical weight is low, the presence of many such rare events may suggest that a distribution with a heavier tail could better capture this behaviour. We leave this possibility to future work.

In contrast, the {\itshape Encyclopaedia Inflationaris} corpus is best described by a slightly different relation:
\begin{equation}
    \label{eq:exponential_4}
    f(r) \sim e^{-r/4}.
\end{equation}
The difference likely stems from several distinctive features of this corpus. All 71 expressions describe the same physical quantity -- the potential energy density of the inflaton field~$\phi$ -- and typically involve only a single variable, reducing the frequency of variable operators. The use of natural units, with all constants except the Planck mass~$M_{\rm P}$ set to unity, simplifies expressions and, as explained below, reduces the occurrence of multiplication and division operators. It also increases the proportion of unary operators, which generally act on dimensionless arguments, as such quantities are more readily formed in natural units. Additionally, the ubiquitous appearance of an energy-density prefactor leads to a higher frequency of the ${\rm pow4}$ operator.
Finally, unlike the largely empirical nature of the other three corpora, the {\itshape Encyclopaedia Inflationaris} includes many hypothetical expressions rooted in speculative high-energy cosmology. This flexibility may allow for a broader variety of operator combinations, contributing to the slower decay rate. Separately, the fact that this corpus deals with much higher energy scales than the others raises the possibility that the decay exponent~$\beta$ could vary with the energy scale.

As in the case of Zipf's law in linguistics, the precise mechanisms driving the observed operator distributions in physical equations are difficult to pinpoint. A natural starting point is dimensional analysis. To produce a ``well-typed'' physical equation, the units associated with variables and constants must match on both sides. This requirement encourages the use of multiplication and division, which combine quantities to maintain dimensional consistency. Numerical coefficients -- often encoding geometrical factors or material properties -- are also common. Low integer powers arise naturally as repeated multiplication, while high powers are rare, likely because they lead to units that are difficult to balance dimensionally.

Notably, transcendental functions such as exponentials, logarithms, and trigonometric functions appear only in the tail of the distribution. This scarcity is again explained by dimensional analysis, as such functions are only permissible when applied to dimensionless arguments -- quantities that must first be carefully constructed from combinations of physical variables and constants.

Beyond dimensional constraints, the structure of physical equations is likely shaped by optimisation pressures, much like the evolution of natural language. Just as Zipfian patterns emerge from the drive for efficient communication, physics benefits from standardised, reusable forms. Cultural factors reinforce this: familiar mathematical constructs are preferred, and frequently used expressions -- such as hyperbolic functions -- are often given compact symbolic representations, streamlining the presentation.

While dimensional analysis and communication optimisation play central roles, the statistical regularities we observe may point to deeper underlying principles -- a kind of ``meta-law'' governing the form of physical laws themselves. Exploring this idea is beyond the scope of the present paper, but we briefly outline a possible approach: one could generate synthetic laws by evolving populations of mathematical expressions (e.g., via genetic programming), constrained only to accurately describe datasets corresponding to physical phenomena. If these artificially generated laws exhibit operator statistics similar to those found in established physical equations, it would suggest that the observed patterns are not merely artifacts of historical convention or communication efficiency, but instead reflect intrinsic structural features of natural laws. Preliminary experiments in this direction support this  interpretation.

\section{Discussion and Applications}
 
\begin{figure}
    \centering
    \includegraphics[width=0.45\textwidth]{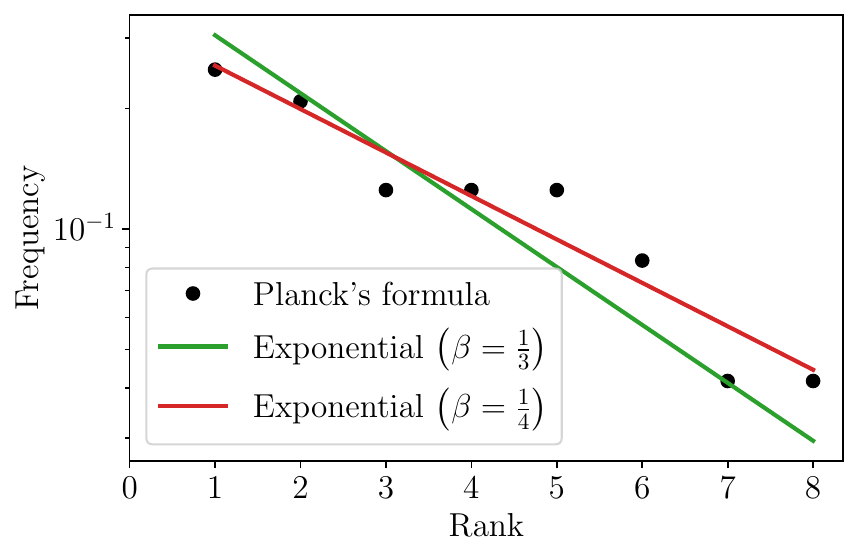}
\caption{\textbf{Operator frequencies in the Planck formula compared to the exponential frequency--rank laws from \cref{eq:exponential_3,eq:exponential_4}}. The plot demonstrates that even individual equations of sufficient complexity exhibit the same statistical patterns found across entire corpora.}   
    \label{Planck}
\end{figure}

Remarkably, the exponential decay observed in operator frequency distributions across entire corpora also holds within individual expressions of sufficiently high complexity. This is illustrated by Planck's formula for the spectral energy density of blackbody radiation:
\begin{equation}
    u_\nu(\nu, T) = \frac{8\pi h\nu^3}{c^3} \frac{1}{e^{\frac{h\nu}{k_B T}} - 1}~.
\end{equation}
When treated as a miniature corpus containing 24 operators, its rank--frequency table reads:
\begin{equation*}
\begin{array}{ll}
(1, 0.25, {\rm mult}) & (5, 0.13, x) \\
(2, 0.21, c) & (6, 0.083, {\rm pow3}) \\
(3, 0.13, a) & (7, 0.042, {\rm exp}) \\
(4, 0.13, {\rm div}) & (8, 0.042, {\rm subtract})~.
\end{array}
\end{equation*}
As shown in \cref{Planck}, the resulting frequency--rank curve closely follows an exponential decay, mirroring the trend found in large-scale corpora. This behaviour is not unique to Planck’s law: across all four corpora, individual expressions of moderate to high complexity (typically containing more than about $10$ operators) consistently exhibit similar distributions. This suggests a striking regularity: physics appears to favour expressions whose internal structure reflects the same statistical pattern seen at the corpus level.
 
This observation has practical implications, particularly for symbolic regression methods aimed at discovering physical laws from data. Building on the statistical regularities identified in real equations, we propose a quantitative measure to distinguish physically meaningful expressions from arbitrary ones. This measure considers two key aspects: (1) how closely an expression's operator frequency follows the exponential frequency--rank relation, and (2) whether the distribution of operator types reflects the empirical pattern observed in physics -- namely, the dominance of variables, constants, multiplication, and division, along with the relative scarcity of unary operators.

Concretely, for any candidate function, one can extract its operator frequency--rank profile and evaluate the probability of such a distribution arising from the empirical statistics observed in physics. This probability may be computed analytically (e.g., assuming Poisson-distributed frequencies) or via simulation (as discussed in the previous section). This could then be implemented as a prior on the function, weighting it according to the extent to which it satisfies the empirical pattern discovered here. This fits naturally into Bayesian or description-length-based model selection methods in symbolic regression~\cite{Bartlett:2022kyi, SRPriors_2023, guimera2020bayesian, 10.1145/3520304.3528899} 
In other model selection schemes, candidate expressions whose operator composition deviates significantly from this prior -- e.g., with too many rare unary operators or too few multiplicative ones -- could be penalised with an appropriately weighted heuristic penalty term, or through a separate objective in a multi-objective optimisation scheme.



In genetic programming approaches, such criteria can guide the construction of the initial population and the selection and mutation strategies during evolution. Seeding the population with expressions sampled from empirical operator distributions could increase the likelihood of generating physically plausible forms early in the search. Similarly, mutation operators can be biased to preserve or reinforce statistical properties aligned with known physical laws.

More broadly, these structural insights may also inform the training or fine-tuning of language models for symbolic reasoning (see, e.g., Refs.~\cite{Biggio_2021,SRPriors_2023}), encouraging them to generate expressions whose statistical and structural features mirror those found in real physics. This approach not only narrows the search space but also enhances the likelihood of identifying physically interpretable mathematical models.

\section*{Acknowledgements}
We~would like to thank Peter Keevash for a fruitful discussion about random binary graphs and Stephen Blundell for pointing out to us a useful reference. We also appreciate the valuable feedback received following the release of the initial preprint version of this manuscript. 
AC's research is supported by a Royal Society Dorothy Hodgkin Fellowship,  grant no.~DHF/R1/231142. DJB was supported by the Simons Collaboration on ``Learning the Universe'' and is supported by Schmidt Sciences through The Eric and Wendy Schmidt AI in Science Fellowship.
HD is supported by a Royal Society University Research Fellowship (grant no. 211046).
PGF acknowledges support from STFC and the Beecroft Trust.

\subsection*{Data and materials availability}
All data supporting the findings of this study, along with the code used to generate random algebraic expressions and perform simulation-based inference, are available at
\url{https://github.com/andrei-const/physics-formulae-patterns}.

\end{document}